RESEARCH ARTICLE                                                                    OPEN ACCESS

# Design of Neural Nonlinear PFC Controller to Control Speed of Autonomous Car

Isam Asaad [1], Bilal Chiha [2]
PhD student [1], Assistant Professor [2]
isam.asaad@gmail.com [1], bilal.chiha@gmail.com [2]
Dept. of Computers and Automatic Control,
Faculty of Mechanical and Electronic Engineering
Tishreen University
Syria

**ABSTRACT**
In this research, we are going to design a neural nonlinear predictive functional controller (PFC) to achieve a reduced fuel consumption for a chosen autonomous car walks according to a supplied speed trajectory on known roads. We used a fitting neural network as a simple tool for modelling the car's engine and control laws needed to calculate the suitable control commands passed to the brakes and gas pedals' actuators. Independent model method and constraints handling are used to provide controller robustness. We used MATLAB Simulink and IPG CarMaker to design and test our PFC controller. The performance of designed PFC controller is compared to the performance of a PI controller which exists within IPG CarMaker simulator.

## I. INTRODUCTION

A self-driving car, also known as a robot car, autonomous car, or driverless car [1], is a vehicle that is capable of sensing its environment and moving with little or no human input [2]. Many decision-makers and practitioners (planners, engineers and analysts) wonder how autonomous vehicles will affect future travelling, roads planning, parking and public transit systems, fuel efficiency and travel costs, shared-use mobility, travel patterns and vehicle design [3], [4]. The full driving task is too complex activity to be fully formalized as sensing-acting robotics systems. One of the approaches to make this job is done using model-based and learning-based methods in order to achieve driver characteristics [5]. We will use a neural network trained to be a nonlinear predictive functional controller (PFC) that is capable of calculating control commands that simulates driver behavior relative to reaching the desired speed, then we will test our controller using IPG CarMaker for Simulink software, which integrates IPG's vehicle dynamics simulation software entirely into the MathWorks' modelling and simulation environment Matlab/Simulink. The highly optimized and robust features of CarMaker were added to the Simulink environment using S-Function implementation and the API functions that are provided by Matlab/Simulink. CarMaker for Simulink is not a loosely coupled co-simulation but a closely linked combination of two best-in-class applications, resulting in a simulation environment that has both good performance and stability [6].

## II. PREDICTIVE FUNCTIONAL CONTROL (PFC)

PFC is one of the most widely used predictive control techniques in industrial applications [7], [8], [9]. PFC, like other MPC methods, uses prediction to select the preferred control action. The prediction depends mainly on the mathematical model of the controlled process, and is evaluated using a performance index, which is a mechanism for selecting the control command that produces the 'best' expected or predicted performance. A key objective of PFC is simplicity of concept and coding. This is a major distinguishing feature compared to more conventional MPC approaches. This simplicity facilitates two advantages [10]:
1- Cost effectiveness (comparable to PID).
2- Ease of implementation (computational and coding requirements are comparable to PID).
The simplicity comes from using limited degrees of freedom in choosing the control command, which will be at most times a choice of a, b, c terms for one of the following forms:

$$u(k)=a; \quad k>=0 \qquad (1)$$
$$u(k)=a+bk; \quad k>=0 \qquad (2)$$
$$u(k)=a+bk+ck^2; k>=0 \qquad (3)$$

Depending on the desired output, we can use equation (1) for constant targets, equation (2) for ramp targets and equation (3) for parabola targets [10]. Since the exact model is known, one can determine the expected steady-state precisely as the following:

$$u_k = u_{ss}, \forall k \geq 0 \qquad (4)$$

$$\lim_{k \to \infty} y_{model}(k) = G_{model}(0) u_{ss} \quad (5)$$

## III. INDEPENDENT MODEL STRUCTURE

Independent model structure is a common tool usually used to ensure unbiased predictions depending on making disturbance estimate or offset term [10], [11]. Figure (1) depicts a typical independent model structure which includes a real process, and an input going into that real process, and what we do is in parallel with a real process; we run a simulation of a system model, and that gives us a model output, so whenever we get an input we always put it into the real process and into the model, however, obviously the model doesn't include offset term, so what we do is measure the real output by measuring the real process output, then comparing it to the model output, which gives us this offset term that tells the difference between real process and modelled system.

This term captures both the actual system disturbance and caters for any error in the modelling parameters, then we can rewrite equation (5) as the following [10]:

$$\lim_{k \to \infty} y_p(k) = G_m(0) u_{ss} + (y_p(k) - y_m(k)) \quad (6)$$

The control law is defined by forcing the predicted asymptotic output to be equal to the desired target 'r' and hence [10]:

$$\lim_{k \to \infty} y_p(k) = G_m(0) u_{ss} + (y_p(k) - y_m(k)) = r \quad (7)$$

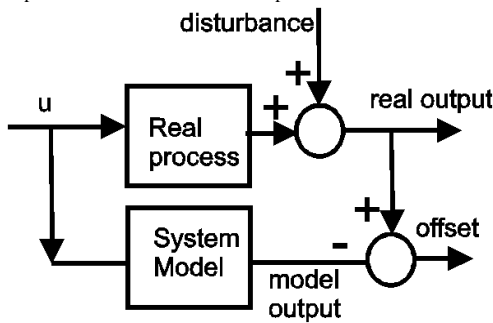

Fig. 1 Typical Independent Model Structure.

Therefore, one can calculate the expected value of the input as [10]:

$$u_{ss} = \frac{r - (y_p(k) - y_m(k))}{G_m(0)} \quad (8)$$

Equation (8) means that the target must be modified to be r-$(y_p(k)-y_m(k))$, where $y_p(k)$ is the measured plant output and $y_m(k)$ is the value of desired setpoint at previous sample. Independent models have the advantage of having the same sensitivity, irrespective of whether it is used with state-space, transfer function or any modelling method.

## IV. CAR ENGINE MATHEMATICAL MODEL AND PFC CONTROLLER DESIGN

Constructing the mathematical model of the controlled system is the first step in designing any predictive controller. The goal of this paper is to design a PFC controller capable of doing a trajectory tracking of a car's speed, and this job is done by controlling the position of the gas and brakes pedals, so our work is to control the performance of the car's engine. In this part, we are going to use the mathematical model of a BMW5 demo car engine included in IPG Car Maker simulator in order to get the control law. Before demonstrating the model used in this paper, we should mention that IPG Car Maker simulator can use models of physical parts made as functional mockup units exported from Dymola and Matlab [12], also this simulator can use models of physical parts and different controllers within Matlab Simulink when using Car Maker for Simulink software. Advanced Mechanical Engineers may fully design their own models, like dual clutch transmission system and engine model made and tested in IPG Car Maker simulator in [12], however, we will use a simple method to get a model of the engine used in any car within IPG Car Maker simulator and its speed controller. Since the full model of the car exists within the simulator, we will design several roads to simulate different scenarios as shown in figure (2), set the car's desired speed setpoint to 80Kmph, and, finally, apply a speed limiter to 50Kmph at zone (1) and a speed limiter to 20Kmph at zone (2) in straight roads (roads 1 to 7).

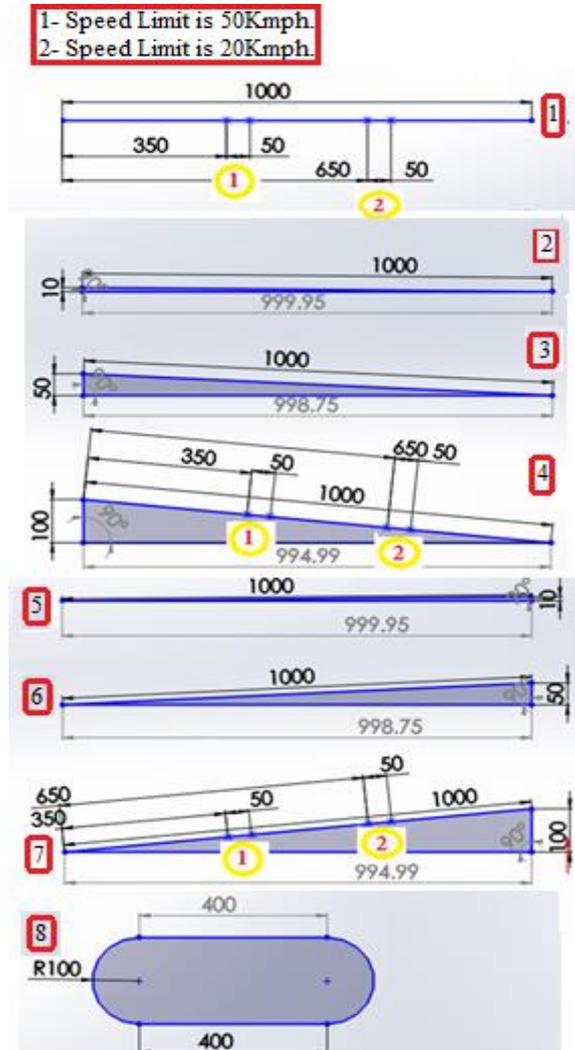

Fig. 2 Roads used to simulate different scenarios.

We make the car move over the shown roads in figure (2), and collect samples of the following resulting values (to be used as controller/model inputs): Desired Car Speed, Current Car Speed, Brakes Pedal actuator's output, Gas Pedal Actuator's output, Gear Number, Car Steer Angle, Car Steer Torque, Engine RPM, Current Car Acceleration, Desired Car Acceleration. Also, we collect samples of the resultant car speed and IPG existing PI control signals of brake and gas pedals as controller/model outputs. Now after collecting these samples, we can train a neural network with the mentioned ten inputs and resultant car speed as one output, as a fitting neural network, as shown in figure (3) to get the car engine model, but what we really need is to obtain the control law by training a neural network which has the mentioned ten inputs and control signals of brakes and gas pedals as two outputs as a fitting neural network, as in figure (4), to get two control signals that leads to get the desired speed given at the first input, taking into consideration the other nine inputs which express the state of the car and its engine. We will use the second neural network (shown in figure (4)) as the PFC car speed controller that achieves the control law used to calculate the control signals for brake and gas pedals. Effect of using this neural network appears as follow:

1- Existed PI controller (made by IPG CarMaker simulation) is reshaped into a neural network.

2- If two or more control commands lead to the same state, neural network will be trained to match the last control command.

3- Every control command (output) is linked to an input state (including a setpoint), which means, there is no more looking for suitable control command as PI do, and this is what make the trained neural network work as a PFC controller able to track a desired trajectory of car speed.

4- If an unknown (untrained) input state appears, neural network will use weights (obtained within training phase) to calculate an approximation of the control command (output).

5- Trained neural network acts as the equations needed to calculate the suitable control commands.

6- Using of trained neural network to express an existed process and its PI controller, relieves the burden of building mathematical model for the car engine and makes it easy to get a 'N step ahead target' PFC controller.

7- The N step ahead target PFC controller is made within neural network training phase. As an example, if we had a training dataset consists of 100 samples, we use samples [1,100-N] as training input set, and samples [N+1,100] as training output set.

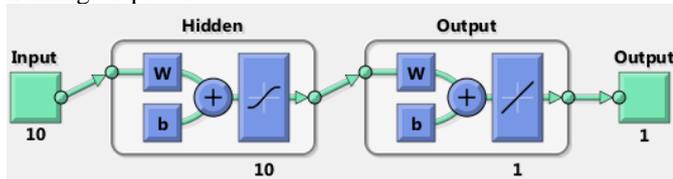

Fig. 3 Trained Fitting Neural Network for car engine model.

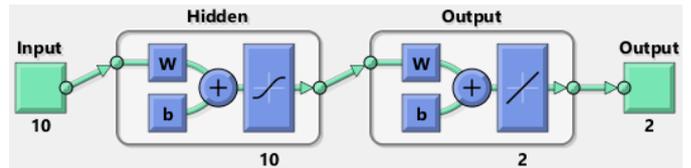

Fig. 4 Trained Fitting Neural Network for car engine PFC speed controller.

Each of the trained fitting neural networks is a two-layer feed-forward network (shown in Figures (3), (4)) with sigmoid hidden neurons and linear output neurons. We used Levenberg-Marquardt backpropagation algorithm [13] achieved by 'trainlm' function in neural network toolbox in Matlab to train the neural networks. The first trained network can be used as a nonlinear mathematical model of the car engine, while we will use the second trained network as a neural nonlinear PFC car speed controller that expresses the two control laws used to choose the suitable control commands passed for brakes and gas pedals' actuators.

The last step of building our predictive controller, after obtaining the nonlinear neural PFC controller, is constraints handling, and this must be done in relation to the amplitude of controller signals. Each of these signals must be within the range [0,1], therefore amplitude limiters are utilized to eliminate any control signal out of this range. The full designed controller is shown in figure (5).

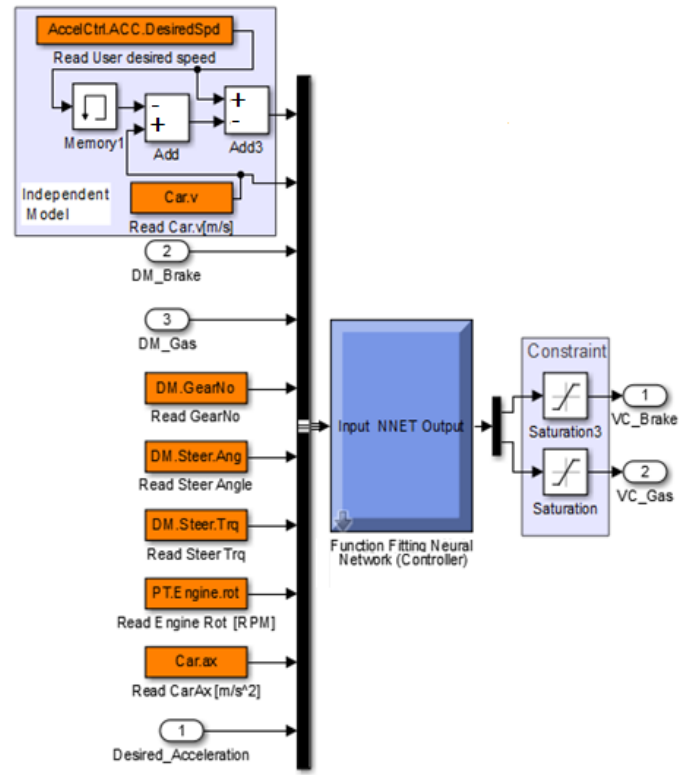

Fig. 5 Designed nonlinear neural PFC controller.

## V. USE OF DESIGNED PFC CONTROLLER WITHIN CAR MODEL

In this paper we used IPG Car Maker software as simulation environment, which can be integrated within Matlab Simulink. Although this integration provides a model of full car within Simulink as S-Functions, IPG CarMaker provides some detailed examples of some parts of the car. We used "AccelCtrl_ACC.mdl" example because it has a detailed Simulink model of PI controller for acceleration/speed of the car works by changing positions of gas and brakes pedals. Now our mission is to replace the IPG PI controller shown in figure (6) with our designed PFC controller (as shown in figure (7)) in order to test it.

VC_Gas and VC_Brake values needed to get the desired car speed, and these two values will always be passed to the S-function of the pedals' actuators.

## VI. RESULTS AND DISCUSSION

In this section, we are going to show the effect of using multiple values of N steps ahead target relative to car fuel consumption. We will choose values 1,4,6,8 and 10 as number

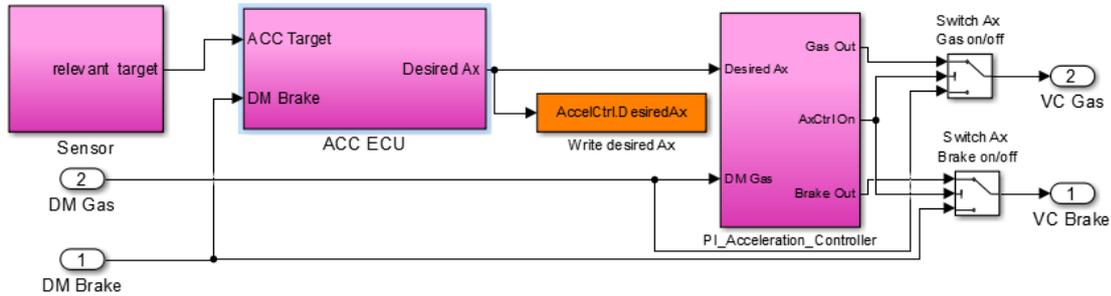

Fig. 6 Placement of PI based IPG car acceleration/speed control system within the car model.

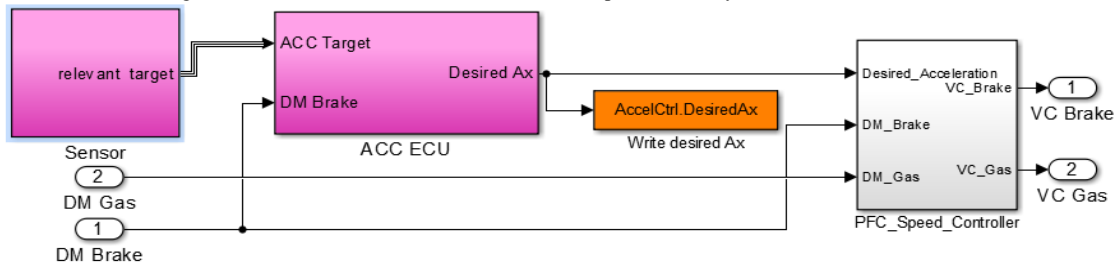

Fig. 7 Placement of designed car speed control system within the car model.

We can see in figure (6) a sensor used to detect objects that may appear within a road like traffic signs, people, cars or any other obstacle. After the sensor appears the acceleration engine control unit (ACC ECU) that calculates the desired car speed and acceleration based on previous sensor's output. Then, a PI acceleration/speed controller appears; this controller calculates the control signals needed to track the acceleration/speed trajectory provided by the acceleration engine control unit. Each of the controller outputs will be passed to a switch, whose output is the vehicle control subsystem control signals (VC_Gas, VC_Brake). These switches pass the controller output to the S-fution of pedals' actuators if the desired car acceleration doesn't equal zero, and pass the previous maneuver control subsystem (DrivMan) values (DM_Gas, DM_Brake) if the desired car acceleration equals zero.

We can see in figure (7) the same sensor and acceleration control unit shown in figure (6), then we can see the designed PFC car speed controller. It is obvious that this controller reads the values shown in figure (5) and always calculates the of steps. After getting trained neural networks, we will organize the results of using these neural networks as nonlinear neural PFC controllers into table (1). This table shows a comparison for using each controller for each of the roads shown in figure (2), relating to car speed setpoint tracking error and fuel consumption enhancement (FCE). Tracking error will be calculated using sum of squared errors (SSE) method. Knowing that the IPG CarMaker simulator gives the absolute fuel consumption of the car, which is the total consumption calculated until the last sample, fuel consumption enhancement is calculated using the following:

FCE = (1 - (Fuel consumption using PFC/Fuel consumption using PI)) * 100   (9)

Now, we are going to use the designed controllers as shown in figures (5) and (7), to make the car walk over the test roads using a target car speed equal to 80 Kmph (80/3.6=22.22mps), taking into consideration speed limiters at zone 1 and zone 2 as shown in figure (2). After that, we will organize our results in table (1).

TABLE I
RESULTS OF USING PI AND DESIGNED CONTROLLERS

| Test Road | | Number of Steps N | | | | | PI |
|---|---|---|---|---|---|---|---|
| | | 1 | 4 | 6 | 8 | 10 | |
| 1 | SSE | 0.0100 | 0.0097 | 0.0096 | 0.0095 | 0.0095 | 0.0100 |
| | FCE% | -0.578 | 1.4869 | 2.1341 | 2.7986 | 3.4253 | 0 |
| 2 | SSE | 0.0097 | 0.0094 | 0.0094 | 0.0093 | 0.0091 | 0.0097 |
|   | FCE% | -0.545 | 1.1609 | 2.5438 | 1.8686 | 2.8568 | 0      |
| 3 | SSE  | 0.0092 | 0.0089 | 0.0087 | 0.0088 | 0.0086 | 0.0092 |
|   | FCE% | -0.528 | 0.8398 | 1.1286 | 0.4081 | 2.2624 | 0      |
| 4 | SSE  | 0.0085 | 0.0082 | 0.0082 | 0.0083 | 0.0082 | 0.0085 |
|   | FCE% | -0.973 | -5.624 | -3.652 | -3.75  | -1.085 | 0      |
| 5 | SSE  | 0.0101 | 0.0098 | 0.0098 | 0.0096 | 0.0096 | 0.0101 |
|   | FCE% | -0.239 | 1.4445 | 2.0421 | 2.3446 | 2.7469 | 0      |
| 6 | SSE  | 0.0106 | 0.0103 | 0.0100 | 0.0101 | 0.0100 | 0.0106 |
|   | FCE% | -0.310 | 1.1836 | 2.2584 | 1.0830 | 1.0950 | 0      |
| 7 | SSE  | 0.0108 | 0.0105 | 0.0103 | 0.0099 | 0.0086 | 0.0108 |
|   | FCE% | -0.046 | 0.7776 | 1.4924 | 1.7960 | 6.9175 | 0      |
| 8 | SSE  | 0.0033 | 0.0032 | 0.0031 | 0.0032 | 0.0030 | 0.0034 |
|   | FCE% | -0.012 | 0.7362 | -3.005 | -11.63 | 2.5138 | 0      |

Table (1) shows the following two results:
1- All designed controllers gave tracking performance comparable to original PI controller.
2- Rising the number of steps N is giving an enhancement in car fuel consumption; but we should say here, that rising the number of steps over 10 did not make any enhancement in fuel consumption.
Since PFC controller with 10 step ahead target gives the best performance within table (1), we are going to show its detailed performance comparison with PI controller made by IPG CarMaker.
We have got the designed neural PFC controller with 10 step ahead target after training the neural network shown in figure (4) for 468 epochs. Then we plotted the linear regression for the training dataset, and got the results shown in figure (8). Figure (8) shows that Levenberg–Marquardt algorithm tried to solve a nonlinear fitting problem.

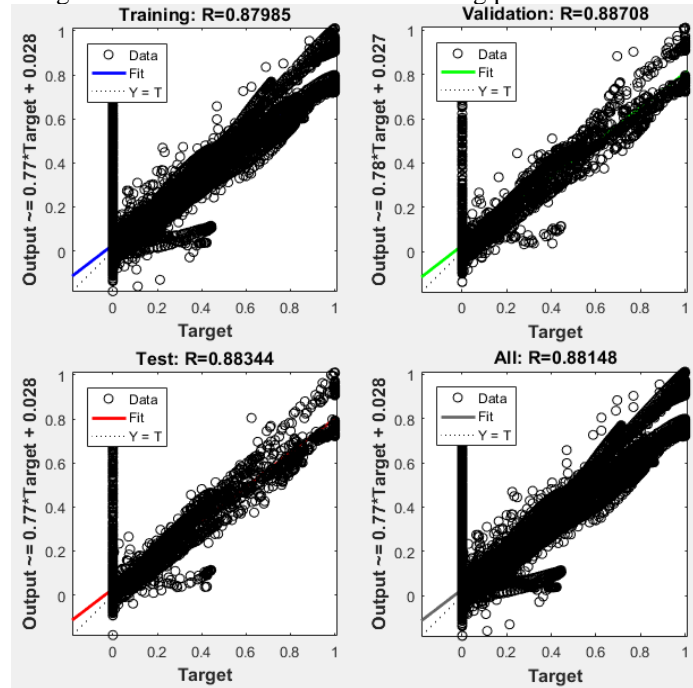

Fig. 8 Linear regression for training dataset targets relative to 10 step ahead target trained neural network outputs.

Finally, the following figures shows detailed performance of the original PI controller and the designed 10 step ahead target PFC controller relating car speed, gear number and fuel consumption magnified 100 times (x100) to make all three results for each road shown in the same figure. Additional figure includes car speed tracking error will be added for each controller while using on each test road. As an example, figure (9) has four parts. Part 1 shows the measured car speed, car gear number and absolute fuel consumption (AFC) magnified 100 times while using IPG CarMaker PI controller to make the car walk over test road (1) (shown in figure (2)). Part 2 shows the difference between measured car speed and desired car speed supplied by the car driver while using the previously mentioned PI controller. Part 3 shows the measured car speed, car gear number and absolute fuel consumption (AFC) magnified 100 times while using the designed 10 step ahead target PFC controller to make the car walk over test road (1). Part 4 shows the difference between measured car speed and desired car speed supplied by the car driver while using the previously mentioned PFC controller.
Figures (10) to (16) shows information relating to the rest of the test roads (shown in figure (2)) in a similar way as shown in figure (9).

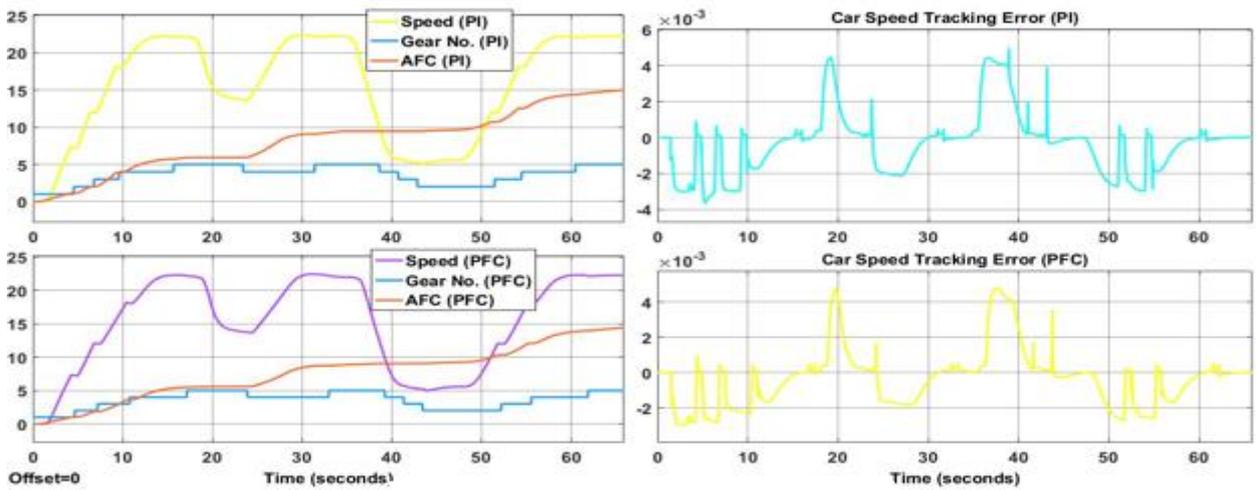

Fig. 9 Comparisons for road 1.

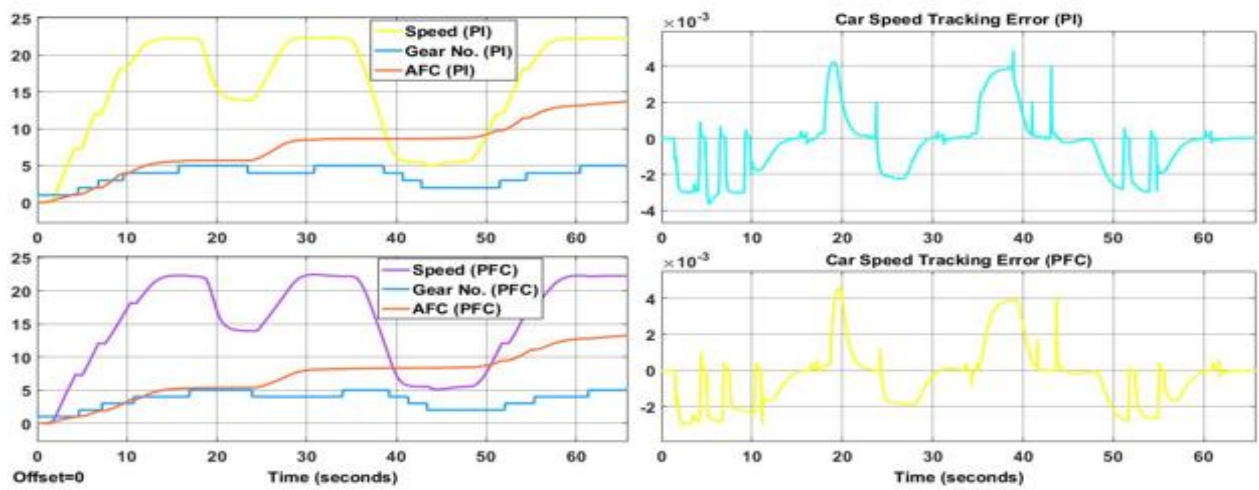

Fig. 10 Comparisons for road 2.

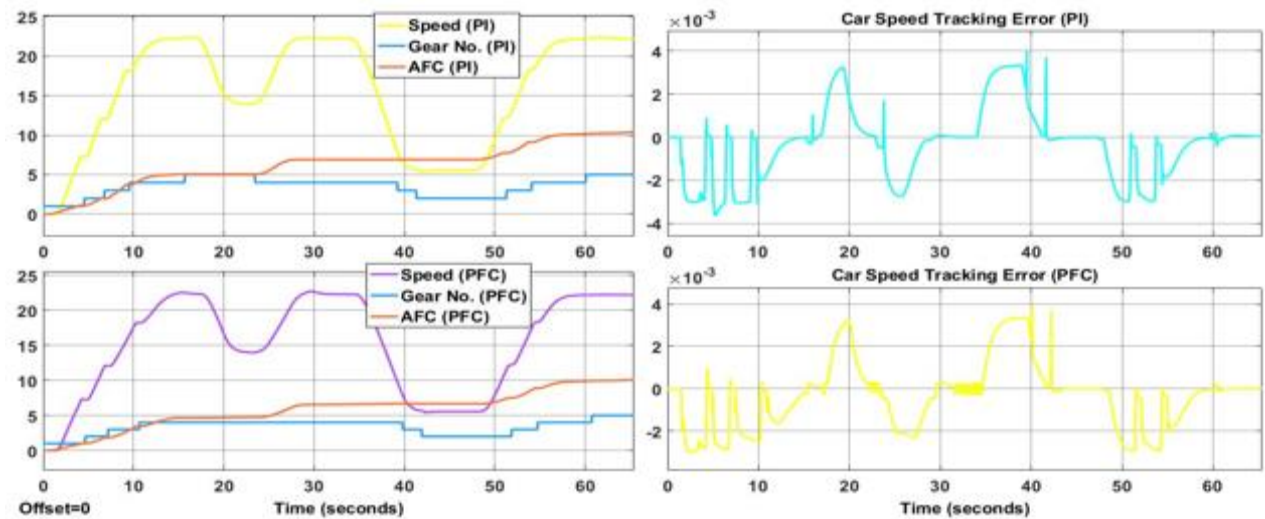

Fig. 11 Comparisons for road 3.

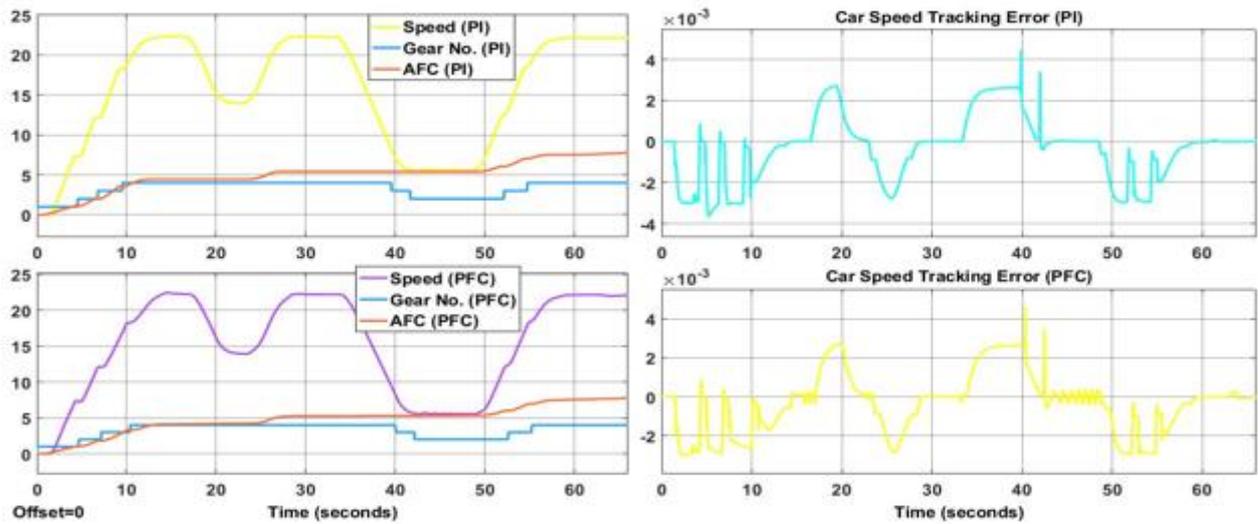

Fig. 12 Comparisons for road 4.

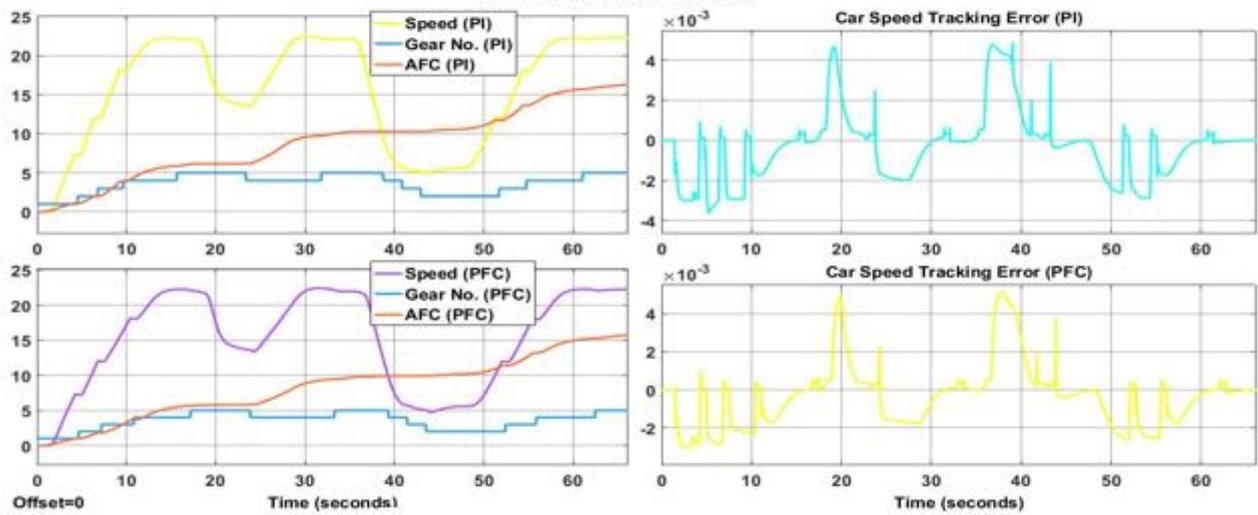

Fig. 13 Comparisons for road 5.

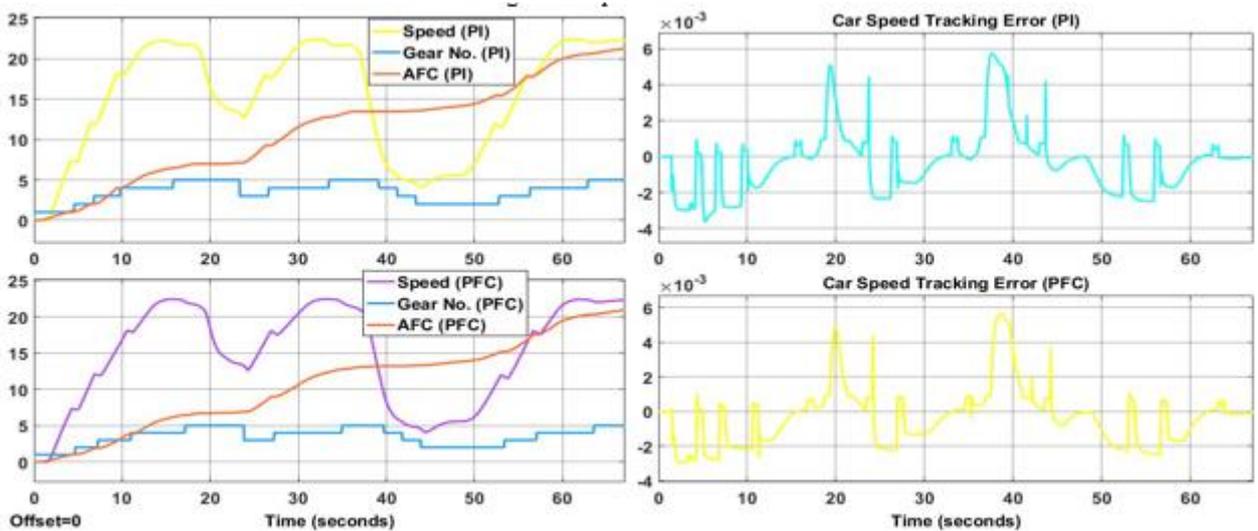

Fig. 14 Comparisons for road 6.

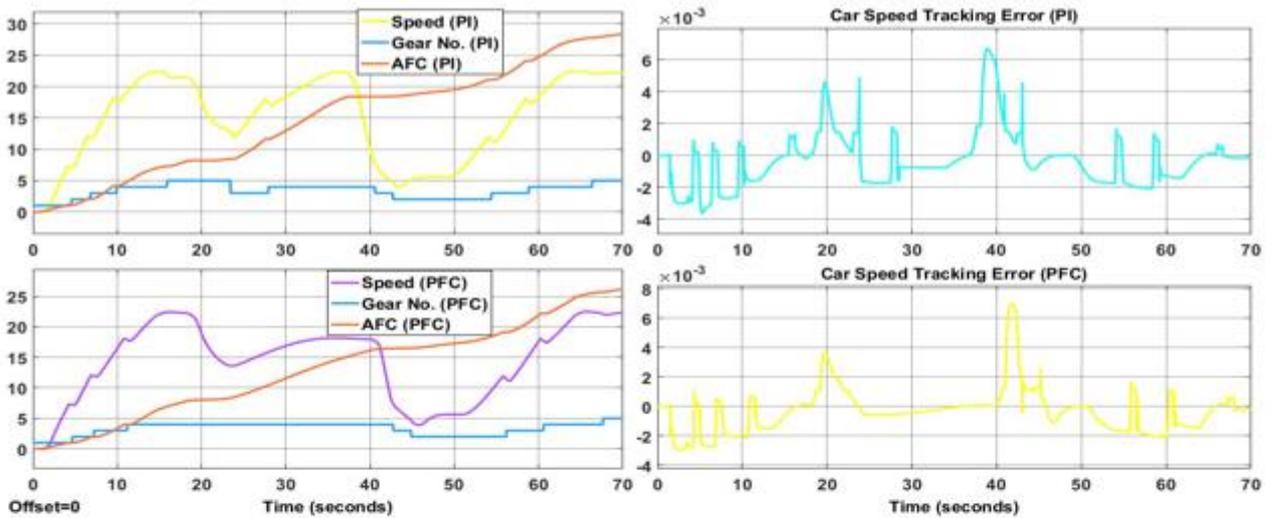

Fig. 15 Comparisons for road 7.

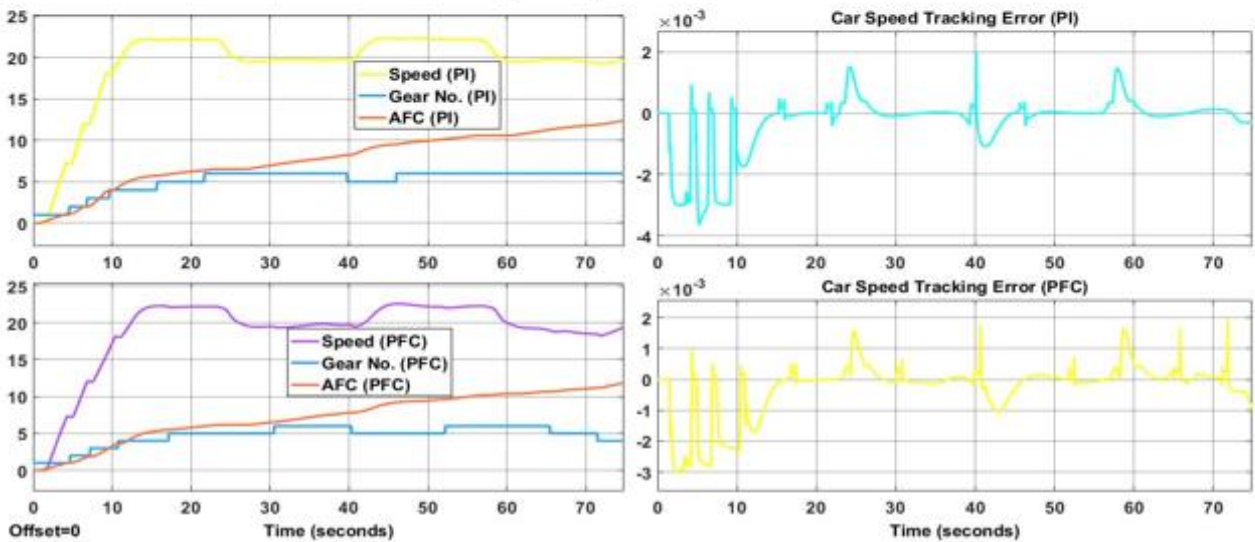

Fig. 16 Comparisons for road 8.

Figures (9) to (16) shows two main results:
1- Designed 10 step ahead target PFC controller could change the behavior of IPG CarMaker Driver. This change affected the trajectory of desired car speed, and gave a trajectory able to reduce the absolute fuel consumption of the car.
2- Fuel consumption reduction had higher values on not inclined ways more than inclined ways.

## V. CONCLUSIONS

In this paper we designed multiple neural nonlinear PFC controllers capable of controlling gas and brakes pedals' position in a self-driving car. We made multiple neural networks to link the engine state with the pedals' position. Inputs of the trained neural network expressed the state of the engine, while outputs expressed the control signals of pedals' actuators. We used the trained neural networks as nonlinear neural PFC controllers with an independent model and one constraint to ensure the controller's performance robustness. Designed 10 step ahead target PFC controller could change the car driver behavior to make an enhancement of fuel consumption compared to PI controller (used by IPG CarMaker simulator) on 7 out of 8 used test ways. This enhancement is clear while applying the designed PFC controller to make the car walk over not-inclined roads more than it is seen when applied over inclined roads.

## ACKNOWLEDGMENT

This work would not have been possible without the support of IPG Automotive. This support included granting a license for 'CarMaker for Simulink' software with a helpful online support. So we should say thanks with gratitude for IPG Automotive team.